\documentclass{acs_template}

\defaultbibliography{bibliography.bib}
\defaultbibliographystyle{achemso}

\begin{document}
\begin{abstract}
Population balance equation (PBE) models have potential to automate many engineering processes with far-reaching implications.
In the pharmaceutical sector, crystallization model-based design can contribute to shortening excessive drug development timelines.
Even so, two major barriers, typical of most transport equations, not just PBEs, have limited this potential.
Notably, the time taken to compute a solution to these models with representative accuracy is frequently limiting. Especially in applications where real-time performance is critical.
Likewise, the model construction process is often tedious and wastes valuable time, owing to the reliance on human expertise to guess constituent models from empirical data.
Hybrid scientific machine learning (SciML) models promise to overcome both barriers through tight integration of neural networks with physical PBE models.
Towards eliminating experimental guesswork, hybrid models facilitate determining physical relationships from data, also known as `discovering physics'. Towards improving efficiency, hybrid models can learn to accelerate numerical models using previous simulation and real-world data.
In this study, we aim to prepare for planned SciML integration through a contemporary implementation of an existing PBE algorithm, one with computational efficiency and differentiability (learnability) at the forefront.
To accomplish this, we utilized JAX, a cutting-edge library for accelerated computing.
We showcase the speed benefits of this modern take on PBE modelling by benchmarking our solver to others we prepared using older, more widespread software.
Primarily among these software tools is the ubiquitous NumPy, where we show JAX achieves up to \num{300}$\times$ relative acceleration in PBE simulations.
This is a significant step towards speeding up slower simulations, even before SciML integration.
Our solver is also fully differentiable, which we demonstrate is the only feasible option for integrating learnable data-driven models at scale.
We show that differentiability can be 40$\times$ faster for optimizing larger models than conventional approaches, which represents the key to neural network integration for further computational processing improvement and physics discovery in later work.
\end{abstract}

\nolinenumbers
\begin{bibunit}

\section{Introduction}
\label{sec-intro}

Transport equations are a powerful tool in designing, optimizing, and controlling a diverse scope of engineering processes, including reaction engineering, adsorption, and carbon capture.
Despite their utility, use of such transport equations often presents two significant challenges in practice. 
The first challenge among these lies in the construction of the models, often limited by the lack of understanding of the underlying physical phenomena, among other reasons.
The second involves the model solution, which, given the complexity of most real systems, is usually non-trivial, requiring significant computational resources and/or time to achieve a realistic solution.

We focus on population balance equations (PBEs), a class of transport equation used to model particle technology (e.g., crystallization, milling), multi-phase reactors (for example, bubble column reactors), and biological engineering (such as fermentation, cell growth/division), among many others \cite{ramkrishnaPopulationBalanceModeling2014}.
They are, in essence, a general mathematical description for the evolution of a particle ensemble.
The challenges with PBEs align with the general challenges in transport modelling we outlined above, and to exemplify these, we specifically discuss PBEs within the context of crystallization processes, commonly encountered in the fine chemical sector.

A crystallization PBE can include terms that describe the crystal population undergoing growth/dissolution, agglomeration, nucleation, and other mechanisms.
These terms are key to the PBE's power and flexibility, but also present a significant challenge during model formulation.
Frequently, these phenomena are not well understood, necessitating the use of empirical equations to fit the behaviour.
From the experimental data, the modeller must often propose several candidate equations that describe the phenomena, and then choose the best fitting one \cite{botschiFeedbackControlSize2019}.
This experimental `guessing-game' depends on heuristics, trial-and-error, and human intervention that incurs additional time and economic costs during the model formulation stage.

Another challenge is the solution of PBEs, which is often difficult to obtain due to their complexity when applied to real-world systems.
The PBEs are typically coupled with additional partial (and/or ordinary) differential equations that model particle interaction with surrounding continuous phase(s).
This results in a system of integro-partial differential equations.
Most often an analytical solution is impossible, which means a numerical one must be calculated instead.
For the purposes of this study, we focus on the finite volume methods (FVM), a family of discretization methods that are widely used in solving PBEs \cite{gunawanHighResolutionAlgorithms2004,maHighresolutionSimulationMultidimensional2002}.
The conservation principle ingrained in the FVM lends itself naturally to PBEs, though it is not without its downsides.
Principal among them is the computational time and resources required to achieve a solution that accurately describes the process.
This problem is exacerbated when the PBE is multidimensional, which arises when multiple internal and/or external coordinates of the particles are being considered, for example when modelling both the size and shape of the crystals and/or coupling fluid dynamics with PBEs.

A concrete example of the impact of the aforementioned PBE modelling challenges can be found in the pharmaceutical industry.
There, many of the crystallization processes being developed for medicines rely on human expertise and \textit{quality-by-testing}.
Transitioning these processes from an art to a science, or \textit{quality-by-design}, has been a long-standing ambition.
The merits of doing so have already been proven for well-behaved systems, but this goal has remained elusive for most systems encountered in reality \cite{cotePerspectivesCurrentState2020}.
The complexity of these real systems makes modelling them especially difficult, which is caused, in part, by the general PBE challenges in model construction and solution.
Overcoming these two challenges will go a long way towards model-based engineering of pharmaceutical processes involving complex crystallization systems.
This is particularly important for an industry where stringent regulation makes the process of developing new medicine exceptionally time-consuming and expensive.

Machine learning (ML) has recently gained popularity in the sciences as a technique to address both challenges we identified.
In crystallization for example, ML has been used to accelerate and improve the accuracy of PBE model solutions, and improve the measurement tools needed to formulate them \cite{xiourasApplicationsArtificialIntelligence2022}.
However, na\"ively replacing existing scientific workflows with ML models often results in poor model interpretability and performance \cite{moseleyPhysicsinformedMachineLearning2022}.
Purely data-driven models discard existing physical knowledge, and therefore need to `re-learn' relevant physics independently.
Worryingly, there is no guarantee that the model will obey physical laws when applied to problems outside its training data.
Furthermore, such black-box models are often hard to interpret and offer no valuable physical insight.

Scientific machine learning (SciML) has newly emerged as a domain to address these concerns, and is undergoing a period of rapid expansion \cite{karniadakisPhysicsinformedMachineLearning2021}.
The basic idea behind SciML is to tightly combine existing scientific workflows and physical knowledge with ML, to improve performance and interpretability.
Many of these hybrid approaches are being developed and gaining popularity, including physics-informed neural networks \cite{raissiPhysicsinformedNeuralNetworks2019}, neural differential equations \cite{kidgerNeuralDifferentialEquations2022}, and other innovative techniques \cite{jumperHighlyAccurateProtein2021}.
These techniques have the potential to significantly improve our approach to PBE modelling, though they are not without their difficulties.
The main one is that, while many of these techniques show promise, their immaturity means their potential is not yet fully understood \cite{moseleyPhysicsinformedMachineLearning2022}.

Typically, the ML components in SciML workflows are trained using gradient descent, and therefore a requirement of most SciML workflows is that the entire workflow is differentiable. 
However, traditional scientific workflows (such as FVM simulations in PBE modelling) are often not developed with differentiability in mind. 
Therefore, a crucial and common first step when developing hybrid SciML workflows is to rebuild existing scientific workflows so that they are differentiable \cite{thuerey2021pbdl,ramsundarDifferentiablePhysicsPosition2021}. 
This allows ML components to be arbitrarily and tightly embedded within traditional algorithms, and the balance between learnability and physical constraints to be finely controlled.

In this work, we make it significantly easier to explore the potential of SciML for PBE modelling by taking this first step; developing a fully differentiable PBE solver. 
Our solver is designed with a SciML-first approach; ML components can be added at any point in the solver, allowing full flexibility in the learnability of the SciML workflow.
Furthermore, the solver is highly efficient, can be run on the CPU or GPU, and can be scaled to handle large amounts of training data.
Another advantage of our approach is that it allows us to pinpoint the computational bottlenecks of our existing algorithms, and any other limitations, before prematurely applying SciML techniques.
We implement our solver using Google's JAX \cite{jax2018github}, a state-of-the-art differentiable programming library in Python which not only makes crafting a differentiable solver and SciML workflows easier, but can also be used to accelerate the performance of any scientific computation.

Our main goal in this work is to showcase a PBE solver using JAX with the following features:
\begin{enumerate}
    \item Differentiability, which will facilitate eventual SciML integration in later work.
    We showcase a proof-of-concept SciML workflow using our solver which allows us to take on the challenge of eliminating the experimental `guessing-game' we mentioned earlier.
    \item Computational efficiency, which aligns with our overarching goal of reducing the computational time and resources required to solve PBEs.
    It will also speed up the process of SciML integration, as ideas can be rapidly prototyped and tested.
\end{enumerate}
Our PBE solver is intended as a strong foundation for SciML integration.
As a consequence, it is also a much improved tool for conventional PBE problems, like forward simulations and parameter estimation, as we will demonstrate herein.

The article is structured as follows. Firstly, in \seref{sec-framework}, we introduce our framework for solving PBEs. Subsequently, in \seref{sec-comp_methods}, we benchmark the proposed JAX-based framework for solving PBEs. In \seref{sec-pest} and \seref{sec-hybrid}, we discuss the differentiability aspects of the proposed JAX-based framework for traditional parameter estimation applications and provide a discussion on future applications to build hybrid models; finally, we conclude with a few closing remarks. 

\section{Population Balance Equation Framework}
\label{sec-framework}

In this section, we introduce our framework for solving PBEs.
In \seref{sec-model}, we present the model that our framework will solve, and the specific numerical methods chosen to solve it.
For this work, we choose to focus on example models within crystallization process modelling, a domain in which PBEs are central.
These examples were chosen to demonstrate the framework's performance, and it is our expectation that the methods used to build it would apply just as well to PBE models outside crystallization, and broadly to any application of the numerical methods we use.
In \seref{sec-methods}, we discuss the solver itself, built in-house using JAX, and e explain its unique features, i.e. differentiability (\seref{sec-auto_diff}) and computational efficiency (\seref{sec-fast}).

\subsection{Model Formulation and Solution}
\label{sec-model}

To demonstrate our solution framework, we first need to choose a model to solve.
On the one hand, our choice should be simple enough to construct and solve, as our scope is limited to constructing a solution framework, and excessive complexity will detract from this focus.
On the other hand, the model should have adequate complexity, so that the computational times taken to solve it are significant enough to be meaningfully compared.
Tangentially, it is important that the solution framework is flexible enough to be extended to more complex models for future work.
It is not enough for the solver to work on the chosen model alone.

Thus, we consider a batch crystallizer where crystals, approximated as cylinders with two characteristic lengths, are subjected to crystal growth \cite{ochsenbeinGrowthRateEstimation2014}.
We assume nucleation and other processes that lead to birth/death of crystals (e.g., agglomeration, breakage) are negligible.
Based on these assumptions, the PBE for a well-mixed batch system can be written as
\begin{equation}
\label{eq-PBE_batch_2d}
    \pdv{f}{t} + \pdv{\left[G_1f\right]}{L_1} + \pdv{\left[G_2f\right]}{L_2}= 0
\end{equation}
where $f$ is the particle size and shape distribution (PSSD), and $L_m$ and $G_m$ are the $m$th characteristic length and the growth rate along that length, respectively.
Two characteristic lengths are considered, giving rise to a 2D PBE, which introduces sufficient complexity for computational solution times to be significant without being excessive.
This 2D PBE is also encountered in real applications of crystallization modelling, as it is the simplest approximation that can take into account both the size and shape of the modelled crystals.

As \equref{eq-PBE_batch_2d} only describes the solid phase, it must be paired with a mass balance describing the corresponding concentration in the liquid phase, which is given by
\begin{equation}
\label{eq-mass_balance}
\dv{c(t)}{t} = -\rho_{\mathrm{c}}k_{\mathrm{v}}\dv{V_\mathrm{c}}{t} = -\rho_{\mathrm{c}}k_{\mathrm{v}}\dv{\left( \int_{0}^{\infty} \int_{0}^{\infty} L_1 L_2^2 f \, \mathrm{d}L_1 \mathrm{d}L_2 \right)}{t}
\end{equation}
where $c$ is the liquid phase concentration, $V_\mathrm{c}$ is the crystal volume, $\rho_\mathrm{c}$ is the crystal density, and $k_\mathrm{v}$ is the shape factor, which accounts for non-cuboidal crystal shapes.

The initial and boundary conditions for \equsref{eq-PBE_batch_2d}{eq-mass_balance} are the following
\begin{align}
    f\left(L_1, L_2,t=0\right) &= f_0\left(L_1, L_2\right) \label{eqn:ic t=0}\\
    f\left(L_1=0, L_2=0,t\right) &= 0  \label{eqn:bc L=0}\\
    f\left(L_1=\infty,L_2,t\right) &=0 \label{eqn:bc L1=inf}\\
    f\left(L_1,L_2=\infty,t\right) &=0 \label{eq-bc L2=inf}\\
    c\left(t=0\right) &= c_0 \label{eqn:ic c=0}
\end{align}
where $f_0$ is the initial distribution, commonly referred to as the seed distribution, and $c_0$ is the initial concentration.

The supersaturation, $S$, is the thermodynamic force driving crystallization, and is defined as the ratio of the solution concentration to the solubility, i.e., $S = c/c^*$.
The solubility is commonly approximated using, for example, exponential or polynomial expressions, depending on what fits the experimental data best.
Similarly, the calculation of the kinetics, which describe the growth of the crystals in our case, can be described using multiple possible empirical equations, including expressions similar to the Arrhenius equation (that capture temperature dependence), and polynomials.
The exact functional forms used for the solubility and the kinetic calculations are not of central importance to our discussion, and thus have been relegated to \appendixref{app-model}.

To solve the PBE model, a high-resolution scheme of the finite volume method (FVM) is used on a fully discretized PSSD \cite{levequeFiniteVolumeMethods2002}.
Additionally, Godunov's dimensional splitting method is used to update the PSSD at each time step along each spatial dimension separately (i.e. twice), using the following expression
\begin{equation}
    \label{eq-highRes_growth}
    \begin{split}
        f^{n+1}_i&=f_i^n - G^n\frac{\Delta t}{\Delta L}\left(f^n_{i} - f^n_{i-1}\right) \\
            &-\frac{1}{2}G^n\frac{\Delta t}{\Delta L}\left(1-G^n\frac{\Delta t}{\Delta L}\right)\left[\phi^n_{i+{\frac{1}{2}}}(f^n_{i+1} -f^n_i) - \phi^n_{i-{\frac{1}{2}}}(f^n_i-f^n_{i-1})\right]
    \end{split}
\end{equation}
where $i$  is the spatial index, $n$ is the time index, $G$ is the growth rate (different for each spatial dimension), and $\phi$ is a flux limiter, which is used to limit numerical oscillations.
More specifically, the van Leer flux limiter is used, because of its stability across a variety of crystallization problems \cite{gunawanHighResolutionAlgorithms2004,maHighresolutionSimulationMultidimensional2002}.
\Equref{eq-highRes_growth} is explicit in time, so to maintain numerical stability, it is constrained by the CFL condition (expressly with a Courant number of 0.9).
The CFL condition is a necessary condition for numerical stability in the explicit time integration of hyperbolic PDEs, and thus is used to determine the stable time step, $\Delta t$, used in \equref{eq-highRes_growth}.
Additional information on the application of high resolution FVM, including the CFL condition equation, is available in \serefSI{si_sec-fvm}.
The mass balance (\equref{eq-mass_balance}) can similarly be discretized to update the state of the liquid phase every time step using
\begin{equation}
    \begin{aligned}
        \label{eq-discrete_mass_balance}
        c^{n+1} = c^n - \rho_\mathrm{c}k_\mathrm{V}&\left[ \int_{0}^{\infty} \int_{0}^{\infty} L_1 L_2^2 f^{n+1}\left(L_1, L_2\right) \, \mathrm{d}L_1 \mathrm{d}L_2\right. \\ 
        -& \left.\int_{0}^{\infty} \int_{0}^{\infty} L_1 L_2^2 f^n\left(L_1, L_2\right) \, \mathrm{d}L_1 \mathrm{d}L_2\right]
    \end{aligned}
\end{equation}
which, once updated, can be used to update the convection rates for the next time step.

To verify the solution obtained by the FVM, we employ the method of moments (MOM), an alternative numerical method.
The MOM solves for average quantities of the distribution, called moments, which are also obtainable by the FVM.
To do this, the MOM recasts the partial differential equation given in \equref{eq-PBE_batch_2d} to a system of ODEs, which can be solved orders of magnitude faster compared to PDEs (like \equref{eq-PBE_batch_2d}), without reliance on a spatially discretized mesh.
Further details on the method of moments are provided in \serefSI{si_sec-mom}.
However, the increased speed and mesh-independence come at a steep cost, as unlike the FVM, the MOM is unable to track the evolution of the whole PSSD.
This means valuable information, like the broadness of the PSSD, is not tracked by the MOM.
Although this downside can be mitigated by increasing the number of moments being tracked or semi-discretization to reconstruct the PSSD, both approaches come with associated downsides.
The aforementioned reasons, as well as the increased robustness of the FVM to more crystallization problems, have made the FVM our numerical method of choice in this study.
However, the computational speed of the MOM, and the availability of efficient and well-tested solvers for the MOM ODEs, make the MOM an ideal validation tool for the FVM solvers we develop in this work.

\subsection{JAX Population Balance Equation Solver}
\label{sec-methods}

Significant developments to open-source languages and libraries in recent years have made them a much better tool at solving the discretized FVM equations presented in \seref{sec-model}.
One such open-source library is JAX \cite{jax2018github}, a cutting-edge Python library designed specifically to meet the computational demands of ML.
While primarily developed for ML applications, JAX is exceptionally suited for most scientific computing requirements, including PBEs.

In this work, we built, from the ground up, a PBE solver framework using JAX.
Two JAX-enabled features central to our study, originally introduced in \seref{sec-intro}, are differentiability and computational acceleration.
We briefly outline their respective importance, and how JAX enables them, in \sesref{sec-auto_diff}{sec-fast}.

\subsubsection{Differentiability}
\label{sec-auto_diff}

The defining feature of our PBE solver is that it is fully differentiable, meaning it can use automatic differentiation (AD) to yield exact derivative values efficiently and flexibly.
AD is especially powerful for vector functions, which map multiple inputs to multiple outputs.
For vector functions, gradients are not scalar, but rather a Jacobian matrix where each element is a single derivative.
The distinguishing feature of AD compared to other differentiation methods, such as numerical, is that it can be used to efficiently calculate an entire column (forward-mode) or entire row (reverse-mode) of a Jacobian at once \cite{bartholomew-biggsAutomaticDifferentiationAlgorithms2000,griewankEvaluatingDerivativesPrinciples2008}.
This makes AD the standard for computing derivatives in the domain of ML, a required operation during training algorithms based on gradient descent.
In fact, backpropagation, the well-known ML training algorithm, is an implementation of reverse-mode AD.
Thus, from an ML perspective, differentiability contributes to the `learnability' of ML models.
This is the reason many kinds of ML models, including neural networks, are programmed to be innately differentiable.
For an extended discussion of the working principle behind AD, along with a simple example, see \serefSI{si_sec-autodiff_basics}.

Likewise, differentiability is an essential requirement when developing hybrid scientific machine learning models. This allows ML components to be inserted into traditional algorithms and trained.
This is the main idea behind differentiable physics \cite{thuerey2021pbdl}.
In practice, this means rebuilding a physical solver using a differentiable library (JAX in our case).
It is worth noting that creating differentiable physical solvers using traditional programming languages with standard libraries, such as Python with NumPy or MATLAB, is either impossible or a non-trivial task.
Such methods have been already been successfully implemented to many models within the physical sciences, where derivatives commonly arise, to enhance derivative calculation.
A few examples include thermodynamics \cite{bellImplementingEquationState2022}, molecular dynamics \cite{schoenholzJAXFrameworkDifferentiable2020}, density functional theory \cite{stierleClassicalDensityFunctional2024}, and synthetic biology \cite{gallupComputationalSyntheticBiology2024}.

The value of using JAX to implement differentiability is in its accessibility.
JAX can independently determine the derivatives of native Python functions, provided a few conditions are met \cite{jax2018github}.
Derivatives of any function can then be returned using a single call of \verb|jax.grad|, which implements reverse-mode AD automatically.
This is highly composable, and derivative evaluations can be easily combined with many other functions and control flows, facilitating complex calculations and computation of higher order derivatives.

As for applications of differentiability to PBEs specifically, there are two of note that will be presented in this work.
The first notable application is to traditional applications that involve derivatives, like parameter estimation.
Another pivotal application is the ability to closely embed and train learnable ML components in the PBE algorithm.
We will use our solver to demonstrate and elaborate on these advantages in \seref{sec-ad_application}.

\subsubsection{Computational Acceleration}
\label{sec-fast}

JAX provides us with computational acceleration in two main ways: software acceleration and hardware acceleration.
The former is achieved using just-in-time (JIT) compilation, which is used to bring the speed benefits to the traditionally interpreted Python, which our solver is written in.
Compiled programming languages are faster than interpreted languages, like Python, because they compile the entire source code to highly efficient machine code before running it, whereas interpreted languages run the code line-by-line at runtime.
The speed-up offered by compiled languages comes at the cost of flexibility and ease of use, as any changes to the source code, however small, require recompilation.
JIT compilation bridges the gap between compilation and interpretation by compiling code just in time for execution.
This avoids a separate compilation step for added flexibility, whilst providing the speed benefits of compilation.
However, the drawback of JIT compilation is that code must be recompiled each time the program is run, which takes time.
Conveniently, this overhead is most often offset by the performance improvements of compilation, especially when the compiled function is used repeatedly.
In JAX, any Python function can be JIT compiled by simply passing it to \verb|jax.jit|, provided the function obeys simple constraints \cite{jax2018github}.

As for hardware acceleration, JAX can leverage GPUs, if installed in the machine where the program is being run.
Using a GPU can significantly speed up calculations that can be performed independently in parallel, which are abundant in the FVM, and has been successfully applied to accelerate PBE simulations previously \cite{szilagyiGraphicalProcessingUnit2016}.
Traditionally, a key issue with GPU programming, commonly performed using NVIDIA's CUDA platform, is that it adds complexity to the development process, making GPU-acceleration a tedious and error-prone addition.
JAX provides a much more accessible approach by automatically optimizing code for execution on GPUs, as long as one is available.
An added benefit is that JAX source code for programs executed on the CPU and GPU are the exact same, which provides much needed flexibility for research and experimentation.
This also gives improved cross-platform compatibility to JAX programs, as JAX searches for a compatible GPU during runtime, and accelerates execution only if found.
These features make JAX extremely efficient on computations involving arrays, of which any scientific computations, including PBE solution, will have plenty.

\begin{figure}[ht!]
    \centering
    \includegraphics{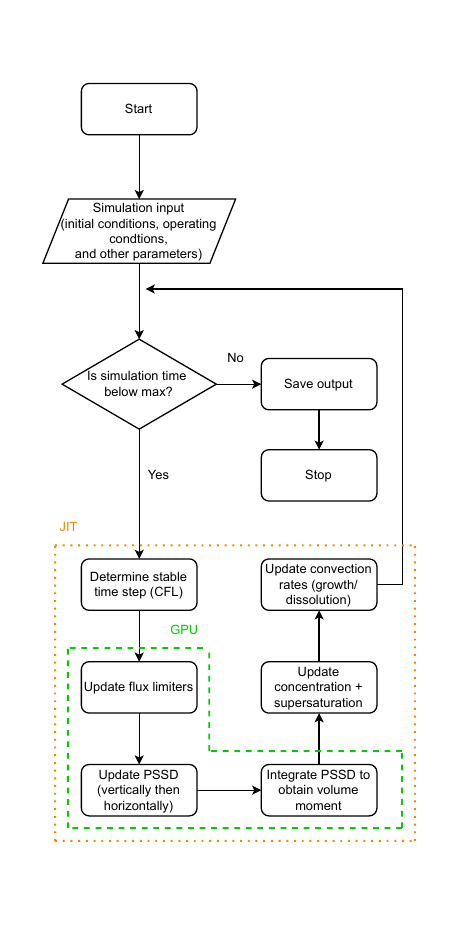}
    \caption[High-level flowchart of the PBE algorithm.]{High-level flowchart of the high-level computation within a PBE algorithm, with operations that would benefit from GPU execution (green dashed region) and/or JIT compilation (orange dotted region) indicated.}
    \label{fig-pbe_algorithm}
\end{figure}

To explain the use of JAX in accelerating PBE solution, we first consider the general FVM algorithm we adopt, shown in \firef{fig-pbe_algorithm}.
The most computationally expensive operation in the algorithm is updating the PSSD, which involves applying \equref{eq-highRes_growth} at each time step, to every element of a large array representing the spatial domain.
Additionally, determining the flux limiters and integrating the PSSD every time step incurs a significant computational expense.
These operations can be repeated hundreds of times per simulation, which presents an opportunity for optimization.
The operations that would benefit the most from being performed on a GPU are enclosed in green dashed lines on \firef{fig-pbe_algorithm}, because they involve the same, relatively simple, calculation performed on each element of a large array.
These identical calculations can be independently performed on thousands of GPU cores simultaneously, which is far faster than a single (or even more) CPU threads.
Also, the operations that would benefit the most from JIT compilation have been highlighted in orange on \firef{fig-pbe_algorithm}. 
This is because these operations include all the calculations that are repeated every time step, which once compiled JIT, can be rapidly executed to significantly reduce the computational time per loop.
JIT compilation also accelerates programs that will run on the GPU, leading to faster parallel execution. 

To our benefit, integrating JAX in Python programs is straightforward, as it has been designed to be a drop-in replacement for NumPy, the established standard in Python array computing.
This means, bar a few notable caveats, many of the programs developed using NumPy can be converted to JAX  without much difficulty, simply by replacing \verb|import numpy| with \verb|import jax.numpy|.
Also, any array operations performed by JAX are inherently differentiable, so they are compatible with JAX's AD functions we mentioned in \seref{sec-auto_diff}.
All the above features have made our PBE solver computationally efficient in its purpose, a fact we explore in \seref{sec-efficiency}.

\section{Computational Solvers Comparison}
\label{sec-comp_methods}

In the present section, we benchmark the performance of our JAX FVM solver framework against the following FVM solvers developed using well-established programming languages and libraries commonly used for scientific computing:
\begin{enumerate}
    \item \textbf{NumPy}: the standard library for scientific computing in Python.
    We use the NumPy FVM solver as a reference for the performance of our other FVM solvers.
    \item \textbf{C++}: a compiled language, which is used to benchmark software acceleration in JAX.
    \item \textbf{NVIDIA's CUDA}: a GPU programming platform provided in C++, which we used to benchmark the hardware acceleration in JAX.
\end{enumerate}
Furthermore, to quantify the impact of using a GPU, we will designate two versions of the JAX solver: JAX (CPU) and JAX (GPU).
The code is the same for both JAX solvers, but access to the GPU is prevented using a software flag for the CPU version (isolating a `software acceleration only' version of the JAX model).
All solvers have been developed in-house, within the scope of this study, and run on the same hardware (AMD Ryzen 3900 12-core CPU and NVIDIA RTX 4090 GPU).
We discuss our procedure for validating our simulations in \seref{sec-verify}, before showcasing benchmarks for our JAX solver in \seref{sec-efficiency}.

\subsection{Simulation Verification}
\label{sec-verify}

It is important before proceeding with any case studies, that the results of the simulations from the different solvers are verified.
To do this, the same simulation is run using the different solvers, and the results are compared to verify they match.
A seeded batch cooling crystallization process is employed as a base case, with crash-cooling simulated at a constant temperature and assuming seeds exhibit a Gaussian distribution.
The parameters used for these base case simulations are summarized in \tabref{tab-default_parameters}.
For the sake of brevity, other required equations and their associated parameters are relegated to \appendixref{app-model}.

\begin{table}[ht!]
\caption{Simulation parameters used by default in all simulations unless stated otherwise}
\label{tab-default_parameters}
\begin{tabular}{@{}lll@{}}
\toprule
Name                                                   & Unit                     & Value                      \\ \midrule
\textit{Initial/Operating conditions}                  &                          &                            \\
Simulation time, $t_\mathrm{max}$                      & [\unit{\hour}]           & \num{1000}                 \\
Temperature, $T$                                       & [\unit{\degreeCelsius}]  & \num{15}                   \\
Seed mass, $m_0$                                       & [\unit{\g}]              & \num{1}                    \\
Initial concentration, $c_0$                           & [\unit{\g\per\kg}]       & \num{8}                    \\
                                                       &                          &                            \\
\textit{Seed parameters (normal distribution)}         &                          &                            \\
Average length, $\overline{L}_1$                       & [\unit{\um}]             & \num{400}                  \\
Average width, $\overline{L}_2$                        & [\unit{\um}]             & \num{250}                  \\
Standard deviation of length, $\sigma_{11}$            & [\unit{\um}]             & \num{30}                   \\
Standard deviation of width, $\sigma_{22}$             & [\unit{\um}]             & \num{30}                   \\
                                                       &                          &                            \\
\textit{Particle properties}                           &                          &                            \\
Crystal density, $\rho_\mathrm{c}$                     & [\unit{\g\per\um\cubed}] & \num{1.11e-12}             \\
Shape factor, $k_\mathrm{v}$                           & [-]                      & $\frac{\pi}{4}$ (cylinder) \\
                                                       &                          &                            \\
\textit{Numerical parameters (spatial discretization)} &                          &                            \\
Maximum length, $L_{1, \mathrm{max}}$                  & [\unit{\um}]             & \num{1200}                 \\
Maximum width, $L_{2, \mathrm{max}}$                   & [\unit{\um}]             & \num{600}                  \\
Bin length, $\Delta L_1$                               & [\unit{\um}]             & \num{1}                    \\
Bin width, $\Delta L_2$                                & [\unit{\um}]             & \num{1}                    \\ \bottomrule
\end{tabular}
\end{table}

\begin{figure}[ht!]
    \centering
    \includegraphics[width=7in]{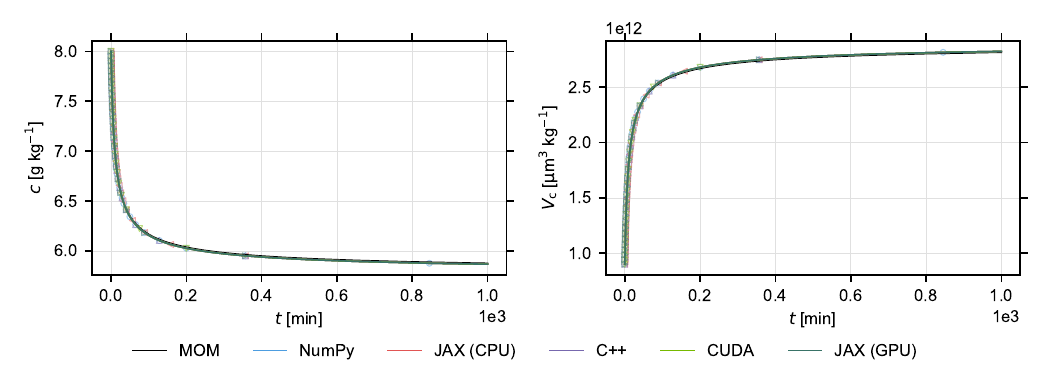}
    \caption[Concentration and total crystal volume profiles comparison.] {Comparison of the concentration (left) and total crystal volume (right) profiles obtained by each FVM solver to those obtained by the MOM.}
    \label{fig-concentration_moment_profiles}
\end{figure}

As mentioned in \seref{sec-model}, we can use the MOM to verify our simulations from each FVM solver used to solve the base case.
To solve the MOM ODEs we use the \verb|scipy.integrate.odeint| ODE solver in Python (with default settings).
The MOM solution yields concentration and crystal volume profiles, which can be used to compare to the profiles obtained by each FVM solver.
We compare these concentration profiles to those obtained from each solver, as shown in \firef{fig-concentration_moment_profiles}.
First, we observe that as the crystal volume increases during growth, the concentration decreases, which satisfies the mass balance (\equref{eq-mass_balance}), as we expect.
We also see that the profiles simulated by each FVM solver and the MOM overlap, which are good indicators that our different FVM solvers are working as intended.

Though it is useful to be able to verify the results of the simulations visually, the large number of simulations completed make this impractical for verifying every simulation. 
Thus, automated tests were implemented to ensure that the results of each simulation presented in this study matches the MOM, and the PSSDs obtained by each solver match each other, in a procedure we outline in \serefSI{si_sec-sim_verification}.
With this automated verification procedure in hand, as well as the ability to inspect our simulations visually on occasion, we can now confidently proceed to quantify the performance of our JAX solver in \seref{sec-efficiency}.

\subsection{Solver Benchmarks}
\label{sec-efficiency}

The efficiency of our JAX solver framework has been assessed by timing a crystal growth simulation, and comparing the time taken by the other FVM solvers to solve the same simulation.
`Wall-clock' time was used as the comparison metric, since it is most relevant with respect to real-world performance.
Computational solution times are averaged over 10 repeats to lessen the impact of random measurement error.
To properly assess the computational efficiency of our JAX solver compared to the alternative solvers we outlined at the start of \seref{sec-comp_methods}, we vary the computational load exerted on the solver using two different methods.

\begin{figure}[H]
    \centering
    \includegraphics{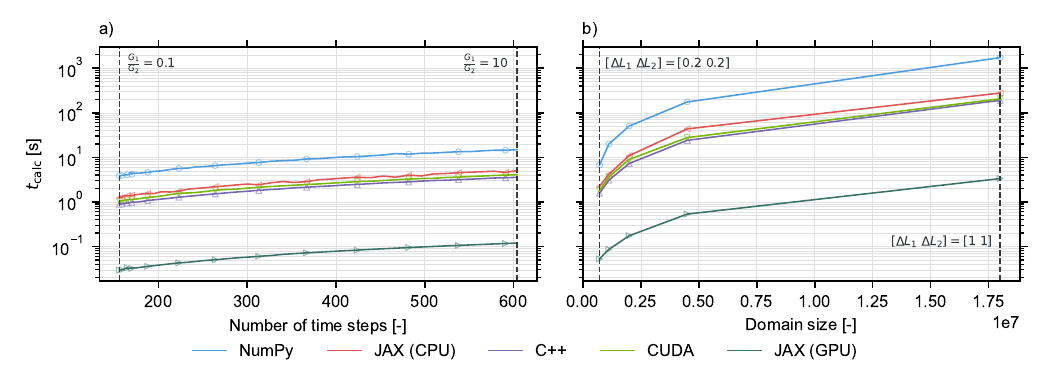}
    \caption[Computation time benchmarks]{Computation time benchmarks for different solvers with varying computational load by varying: a) the number of time steps, and b) the spatial domain size. Both extremes for each varied parameter are marked by vertical dashed lines. The CUDA and JAX (GPU) solvers have access to a GPU, while the remaining solvers are CPU-based.}
    \label{fig-comparisons}
\end{figure}

The first of these methods is varying the number of time steps, which translates to more iterations of the loop shown in \firef{fig-pbe_algorithm}.
To vary the number of time steps we vary the relative growth rates from their original values (\tabref{tab-additional_parameters}), while fixing the rest of the parameters.
More specifically, we vary the ratio $\frac{G_{1}}{G_{2}}$ from its original value by a factor of \num{0.1} to \num{10}.
These growth rate ratios are dependent on the crystalline compounds making up the system being modelled, such as different crystal polymorphs.
The results are shown in panel a) of \firef{fig-comparisons}, which plots the computational calculation time, $t_\mathrm{calc}$, in logarithmic scale against the number of time steps.
First, for all solvers, we observe that the calculation time increases noticeably with increasing number of time steps, as expected.
Secondly, we observe that the NumPy solver is the slowest, which is also expected as NumPy is severely limited by the Python interpreter, hence its use as a reference.
Unsurprisingly, the introduction of a compilation step by the C++ solver has led to a significant decrease in calculation time compared to the NumPy solver, which is because compiled machine code is faster than interpreting code line by line.
Finally, we observe that the JAX solver, even when restricted to CPU use, achieves similar speeds as the C++ solver, despite being based in Python.
The reason is the JAX solver, even without access to a GPU, is still able to leverage JIT compilation on the CPU to achieve software acceleration.
This result is made significant by the fact that the solver based in C++ was much more difficult to develop, and less flexible to use, than the JAX solver.
On average, with access to only a CPU, the JAX solver is able to achieve a \num{3}$\times$ speed-up over the NumPy solver.

As for the GPU solvers (CUDA and JAX) shown in panel a) in \firef{fig-comparisons}, our first observation is that the CUDA solver is slightly slower than the C++ solver, which demonstrates the difficulty of manual GPU programming.
The CUDA solver took by far the most time to develop, but this development time did not translate to any computational efficiency gain.
We were able to identify memory transfer between the CPU and GPU as the bottleneck, but optimizing will incur additional development time.
It is known that CUDA has potential to speed up PBE solution \cite{szilagyiGraphicalProcessingUnit2016}, but this potential is not properly explored in the scope of our study.
In contrast, the JAX solver, once allowed GPU access, is much faster than our CUDA solver despite taking a fraction of the development time.
While theoretically possible that the CUDA solver could be made much faster through time-consuming optimization, we believe that the accessible solution offered by JAX is best for most scientists and engineers such as we.
Comparing the JAX solver to the NumPy solver, we got an average speed-up of a factor of \num{130}$\times$ with varying time steps.

The next method we investigate in varying computational load on the solver is a varying spatial domain size, which refers to the number of bins, or elements, that represent the spatial domain.
The bin size, also known as the mesh size, is an important numerical parameter that dictates the accuracy of the PBE simulation.
Ideally, the bin size should be as small as possible for improved accuracy, though this comes at the cost of increased computational time and memory usage.
This is because the number of bins is inversely proportional to the square of the bin size (in the 2D case), which exponentially increases the number of calculations required each time step as the bin size decreases.
The reason for this is \equref{eq-highRes_growth} is applied to each bin independently, so more bins means more equations to solve.
Relating this back to the flow chart in \firef{fig-pbe_algorithm}, this represents increasing the computational cost of the steps highlighted in green.
In addition to all the above, an increased bin density also requires an increased number of time steps, which translates to more iterations of these heavy calculations.
Once again, the simulation parameters outlined in \tabref{tab-default_parameters} are used, with the bin area (the product of the bin sizes) varied uniformly from \qtyrange{0.04}{1}{\um\squared}.
The results of this bin size variation are shown in panel b) of \firef{fig-comparisons}, which similar to panel a), plots the calculation time in log-scale against the spatial domain size.
Our immediate observation is that the calculation time increases much more significantly with the domain size compared to the number of time steps alone.
This is because there is not only more time steps, but the calculations at each time step are more intensive as the domain size increases.
Otherwise, the trends relating the different solvers are similar to panel a).
The biggest difference is that now the average speed of the JAX (GPU) solver is, on average, \num{300}$\times$ faster than NumPy.
This demonstrates that our JAX solver is consistently fast and scales well to increased computational loads.

Out of curiosity, we also compared the speed of an existing PBE solver developed by the senior author and reported in 2019 \cite{botschiFeedbackControlSize2019}.
That solver was created in MATLAB and optimized with all the features available at the time, including MEX C++ compilation.
The solver was run using its default bin density of \numproduct{360x120}, which is a much lighter computational load than the minimum bin density we have subjected our solver to (\numproduct{1000x500}).
The time taken by the MATLAB solver to complete 234 time steps was \qty{1.2}{\s}.
Even comparing this time with our slowest simulation result (a bin density of \numproduct{6000x3000} which required \num{1339} time steps), our JAX (GPU) solver is still almost three times as fast.
The comparison to the MATLAB solver isn't rigorous by any means, but suffices as a qualitative measure of the progress the JAX solver has made relative to a recent alternative.

\section{Differentiable Physics Applications}
\label{sec-ad_application}

Within this section, we address how the differentiability of our PBE solver can be applied.
First, in \seref{sec-pest}, we present how differentiable physics can be used in traditional workflows by considering a parameter estimation example.
Then, in \seref{sec-hybrid}, based on our findings from the example, we conclude this section with a discussion on new applications that differentiability facilitates, which we will explore in upcoming studies.

\subsection{Traditional Physical Applications}
\label{sec-pest}
In \seref{sec-auto_diff}, we introduced the concept of differentiable physics, where pre-existing physical solver algorithms are augmented with the ability to apply automatic differentiation (AD).
We briefly mentioned that differentiable physics also has its applications to traditional physical applications, which we will demonstrate in this section by way of example.
In this study, we choose parameter estimation as that example, a common inverse problem in science and engineering.

The objective of parameter estimation is to estimate the true values of parameters characterizing empirical models, which can be stated mathematically as
\begin{equation}
    \label{eq-loss}
    \min_{\boldsymbol{\theta}} \;  L( \mathbf{y}, \mathbf{\tilde{y}}(\boldsymbol{\theta}))
\end{equation}
where $L$ is an objective or loss function to be minimized, which quantifies the deviation between the true data, $\mathbf{y}$, and the model's prediction of that data, $\mathbf{\tilde{y}}$.
The model prediction is a function of the parameters, $\boldsymbol{\theta}$, which must be estimated from experimental data.
To estimate the parameters, gradient-based optimization algorithms are often used, which require the gradient of the loss function with respect to each of the parameters in matrix form, also known as the Jacobian of the loss function.
These algorithms are often available for out-of-the-box use, such as MATLAB's \verb|fmincon| and Python's \verb|scipy.optimize.minimize| (SciPy).
Since the loss function maps a vector of parameters to a scalar (for single-objective optimization), its Jacobian is a row vector of the form $[\pdv{L}{\theta_1}, \ \cdots, \ \pdv{L}{\theta_k}]$ for $k$ parameters.
AD is well-suited to the calculation of this single-row Jacobian such as this, as a single pass of reverse-mode can be used to calculate it (see \seref{sec-auto_diff}).

To assess the application of AD to population balance equations, we continue with the crystallization model introduced in \seref{sec-verify}.
More specifically, we focus on the parameter estimation of the growth rate, which is typically performed after a suitable candidate model has been proposed.
In our example, the proposed growth rate model is the following $n$-th degree polynomial
\begin{equation}
    \label{eq-poly_growth_rate}
    \tilde{G}_m(S) = a_1 (S-1) + a_2 (S-1)^2 + \ldots + a_k (S-1)^k
\end{equation}
where $a_1, a_2, \ldots, a_n$ are the growth rate parameters to be estimated (analogous to $\boldsymbol{\theta}$ from \equref{eq-loss}).
The reason we chose this expression, despite its apparent impracticality, was because the number of parameters to estimate can easily be increased (by increasing the number of polynomial terms). However, we acknowledge that in real-life scenarios, this expression will hardly ever be used.

The data and associated procedure for simulating the experimental data needed to fit the growth parameters can be found in \appendixref{app-sim_experiments}.
Once experimental data had been generated, we used our differentiable solver, demonstrated in \seref{sec-efficiency}, to simulate the same data, but based on the prediction of the growth rate model (equivalent to $\mathbf{\tilde{y}}$ from above).
The optimization algorithm we chose is Adam, due to its popularity in the machine learning field for features including adaptive step sizes \cite{kingmaAdamMethodStochastic2017}.
Other details on our optimization procedure can be found in \appendixref{app-pest}.
As Adam is a gradient-based method, it requires the gradient of the loss function with respect to each of the parameters, which is a row Jacobian that we must supply.
Since our differentiable solver is JAX-based, we can easily implement reverse-mode AD to estimate the gradient of the loss function with respect to each parameter, as we discussed in \seref{sec-auto_diff}.
To quantify the performance of JAX's AD algorithm (jax-AD) we compare to two forward-difference numerical differentiation (ND) algorithms: an ND algorithm accelerated using JAX features (jax-ND) and a na\"ive ND algorithm with no acceleration (slow-ND).
Note that common numerical differentiation algorithms, such as the one leveraged by MATLAB's \verb|fmincon|, should theoretically have a level of performance that is in-between the two ND algorithms (more likely closer to the latter, i.e. the na\"ive ND algorithm), but definitively proving this is out-of-scope of this work. 
The comparison of AD with ND is appropriate, as ND is the default choice for gradient estimation in applied PBE modelling employing gradient-based parameter estimation and optimization techniques \cite{ochsenbeinGrowthRateEstimation2014, botschiFeedbackControlSize2019, salvatoriExperimentalCharacterizationMathematical2018}.

\begin{figure}[ht!]
    \centering
    \includegraphics{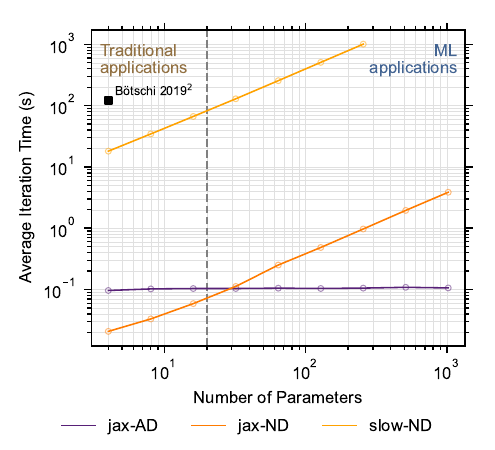}
    \caption{Comparison of the average iteration time during parameter estimation with increasing number of parameters for distinct differentiation algorithms: jax-AD, jax-ND, and slow-ND. The single marker indicates the time taken by the Bötschi 2019 solver \cite{botschiFeedbackControlSize2019} to update 4 parameters over a single iteration. Two regions have been highlighted, which indicate the number of parameters that are normally optimized for traditional (number of parameters~$\lessapprox 20$) and machine learning (number of parameters~$\gtrapprox 20$) applications.}
    \label{fig-pest}
\end{figure}

The results of the parameter-estimation study using the three different gradient calculation algorithms are presented in \firef{fig-pest}, where the average time taken for the optimization algorithm to complete a single iteration (an average over 100 iterations) using the three gradient calculation algorithms is shown.
First, we draw attention to the two ND algorithms, where we observe that the jax-ND algorithm is orders of magnitude faster than the slow-ND algorithm.
This is due to the numerous acceleration features that JAX offers, not only software and hardware acceleration (as demonstrated in \seref{sec-efficiency}), but also the ability to efficiently vectorize the calculation of a Jacobian element with relative ease.
This vectorization is conceptually similar to vectorized operations in NumPy and MATLAB, and is achieved by executing the same numerical gradient calculation to each element of the row Jacobian in concurrent batches on the GPU.

Moving on to the comparison of the jax-AD algorithm with the jax-ND algorithm, our initial observation is that the jax-ND algorithm is faster than the jax-AD algorithm for estimating only a few parameters.
While this advantage is much more pronounced for a small number of parameters, it quickly diminishes with an increasing number of parameters.
One of the factors contributing to this finding is a well-known disadvantage of reverse-mode AD \cite{griewankFundamentalsForwardReverse2008}.
While reverse-mode AD theoretically requires fewer operations to compute a row Jacobian compared to ND, this computational efficiency comes at the cost of increased memory usage.
The memory issue was especially prominent in our case because we used a GPU, which typically has much less memory available when compared to CPUs.
To mitigate the memory usage, we needed to use checkpointing, which is a reverse-mode AD technique that trades computational speed for memory.
Ultimately, ND is more efficient in practice for traditional physical applications, where the number of parameters to optimize is often small (less than 20).
However, it is worth noting that AD is exact (to machine precision), while the accuracy of ND depends on numerical factors including the step size and the numerical scheme used.
This means that it can be more convenient to use AD anyway, especially in applications where the choice of the aforementioned numerical parameters is not straightforward.

Finally, the comparison of jax-AD and jax-ND with the slow-ND algorithm reveals that, while the performance of the both ND algorithms deteriorates exponentially with increasing number of estimated parameters, the performance of AD remains relatively constant no matter how many parameters are to be estimated.
For \num{1000} parameters specifically, jax-AD is \num{40}$\times$ faster than jax-ND.
This result is significant, as it demonstrates that AD is the only feasible choice for optimization problems with a large number of parameters, which are common in machine learning applications.
The weights and biases of a neural network, for example, are analogous to the parameters we estimate in physical application, except they often number in the hundreds or thousands.
In fact, the backpropagation algorithm that is ubiquitous in machine learning is an implementation of reverse-mode AD, which is only possible because neural networks are inherently differentiable.
Thus, the differentiabilty of our solver means we can embed and train a neural network within our solver, which we discuss planned applications of in \seref{sec-hybrid}.

Similar to the discussion presented in \seref{sec-efficiency}, we compare the performance of the optimization algorithm to the one reported by the senior author in 2019 \cite{botschiFeedbackControlSize2019}.
There, MATLAB's \verb|fmincon| was used as the optimization algorithm, with the senior author's MATLAB MEX PBE solver at the centre.
For one of the proposed models from that paper, four dissolution parameters were estimated using data from seven experiments.
We ran their optimization code using the same machine we have used throughout this work.
For added context, as we mentioned in the conclusion of \seref{sec-efficiency}, the computational load exerted on the MATLAB solver during optimization is much lighter than the one our solver handles (due to the smaller bin density used).
The time taken by the MATLAB optimizer to update the parameters at each iteration is about \qty{120.3}{\s}.
This result has also been annotated on \firef{fig-pest} to compare to our own optimization algorithms, which shows that jax-AD and jax-ND are roughly \num{1200}$\times$ and \num{6000}$\times$ faster at estimating 4 parameters, respectively.
The implications of this finding are significant, as despite the increased accuracy, our optimization performance is able to approach real-time optimization (for a few parameters).
While we do not intend the comparison to the MATLAB optimizer to be rigorous, it gives us an idea of the optimization performance improvements our solver has made compared to recent efforts.

\subsection{Hybrid Model Applications}
\label{sec-hybrid}

Throughout this work, we tried to emphasize the general applicability of the techniques we used, not only to other PBE applications, but to any field involving the solution of partial differential equations, more specifically transport equations, across science and engineering. The availability of a modern, efficient, and differentiable model opens up new avenues to incorporate machine learning tools into a physical solver, which we will explore in our upcoming studies. Here, we will discuss a particular class of SciML techniques, known as hybrid models.

Hybrid models replace only part of a physical solver with a learnable component, such as a neural network.
This gives us the freedom to approximate only the parts of our solver where we have incomplete knowledge, or those that are computationally expensive.
When applied in this manner, the neural network is physically constrained by the fixed solver surrounding it, while maintaining its data-driven flexibility.
This means that even the worst trained hybrid models will not be able to violate these fixed physical laws, which is an advantage over methods that attempt to approximate the physical solver's behaviour wholesale, i.e. through purely data driven approaches.
The potential benefits of this strategy, aside from flexibility and performance improvements to the physical solver itself, include the fact that the neural network can be trained more easily with less data \cite{moseleyPhysicsinformedMachineLearning2022}.
These hybrid techniques can only be properly explored by the use of a differentiable solver, as without one, the size and flexibility of the embedded neural network is severely limited.
This is because traditional numerical differentiation does not scale well to models of a large size.
We have shown this to be true in \seref{sec-pest}, and it has also been shown in the literature for neural networks trained with a numerical gradient-based optimizer \cite{narayananHybridModelsSimulation2021}.

\begin{figure}[ht!]
    \centering
    \includegraphics{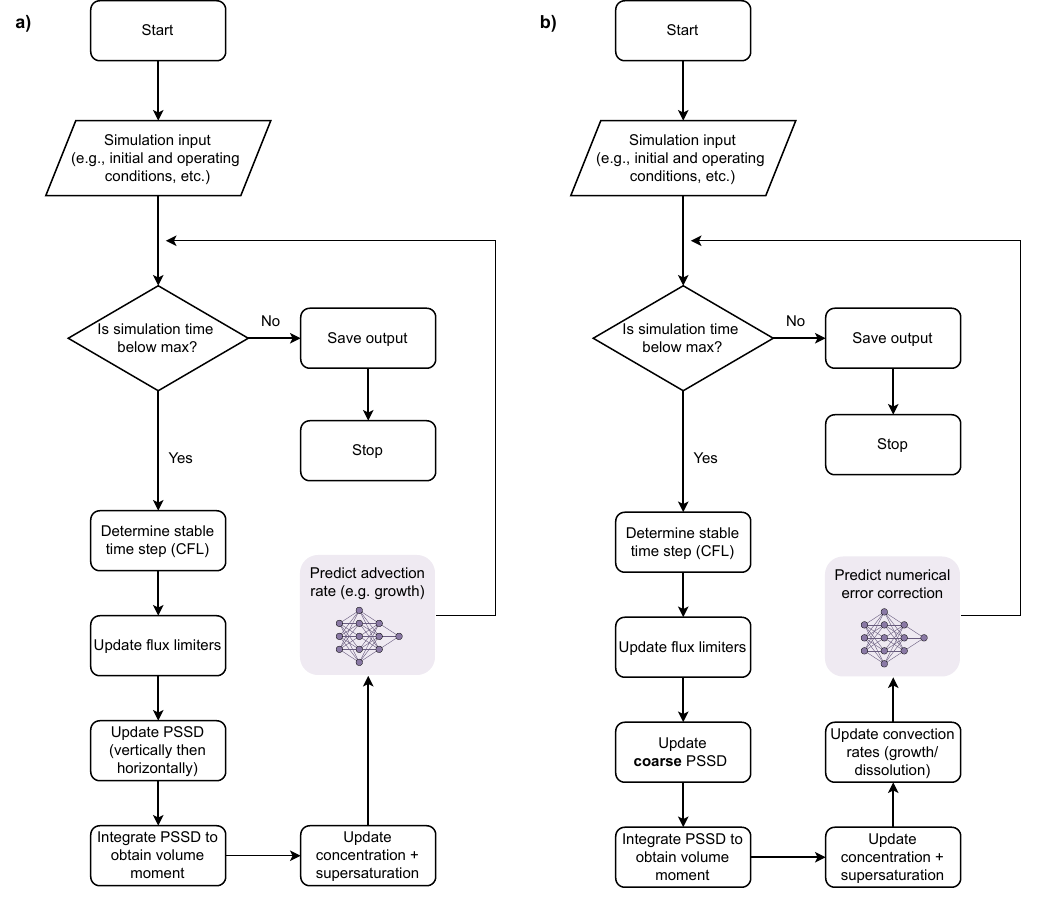}
    \caption{Two conceptual illustrations of potential hybrid models applied to the algorithm from \firef{fig-pbe_algorithm}: a) Empirical hybrid model, where a neural network replaces the calculation of the advection rate. b) In-the-loop model, where a neural network estimates the discretization error of the solver at each iteration of the loop.}
    \label{fig-hybrid}
\end{figure}

Since we have already created a differentiable solver for PBEs, we now discuss potential hybrid PBE models.
To continue the crystallization example from \seref{sec-pest}, instead of a proposed growth rate polynomial model, we can replace the growth rate with a neural network prediction.
A depiction of this empirical hybrid model is shown in \firef{fig-hybrid} a), which indicates where the neural network will be embedded, and the fixed algorithm surrounding it.
The weights and biases of the neural network would then be estimated during training in the exact same way as we have estimated the parameters in the polynomial growth rate model (\equref{eq-poly_growth_rate}).
Since this neural network is theoretically a universal approximator, it promises a more flexible approach to determining the growth rate.
By using a hybrid model in this context, we can ideally target the `best of both worlds': the certitude of concrete physical knowledge, including the mass and number balances, present in the rest of our solver, and eliminating the empirical guessing game with a black-box only where necessary.
The field of crystallization has already seen promising results with hybrid models from the literature, which encourages us to explore this approach further \cite{sitapureIntroducingHybridModeling2023}.
Additionally, with tools like symbolic regression \cite{cranmerInterpretableMachineLearning2023}, it is possible to decompose the neural network into an interpretable equation.
Combining this tool with hybrid models, we envision an automated pipeline that is capable of discovering the empirical equations for phenomena underpinning PBEs, eliminating the need for human guesswork during model development.
However, one of the challenges we anticipate is that, even with the surrounding PBE solver in place, the neural network is not guaranteed to converge to a physically meaningful expression.
This challenge must be resolved, as settling for a data-driven model that essentially acts as a fitting tool, without discovery of the physical phenomena involved, is dubious at best for industrial adoption and gaining research insight.

Additionally, the use of hybrid models is not limited to the prediction of empirical models for phenomena, but can also be used to enhance the numerical performance of the solver itself.
For example, a promising family of techniques are the so-called `in-the-loop' methods \cite{umSolverLoopLearningDifferentiable2020}, which exploit the iterative nature of algorithms for solving partial differential equations, which includes many PBE solvers.
Similar to the empirical hybrid models we discussed above, the idea is to embed a neural network within the solver, but this time to estimate the discretization error of the solver itself, rather than approximate some physical phenomena.
A visualization of the in-the-loop model is exhibited in \firef{fig-hybrid}~b), which illustrates where it can be inserted to offset numerical error. 
This in-the-loop model, once trained, can be used to obtain high-resolution accuracy, with the speed of a faster coarse model.
The difference in solution time between a coarse and fine grid simulation can approximately be \num{100}$\times$ in some cases (see \firef{fig-comparisons}), which is a significant speed-up that can be achieved with a hybrid model.
This is not to mention the speed-up we already achieve without a neural network in JAX (up to \num{300}$\times$), which together with hybrid models, can be used to progress towards real-time optimization and control of industrial processes.

\section{Concluding Remarks}
\label{sec-conclusion}

In this study, we presented a PBE solver built from the ground up using JAX, an accelerated scientific computing library, and applied it to a variety of common crystallization problems. The key outcomes from this work can be summarized as follows:
\begin{itemize}
    \item one can achieve a speed-up of 300$\times$ relative to conventional PBE solvers, which we showed is scalable and consistent across different numerical and operating conditions
    \item one can have a fully differentiable PBE solver facilitated by automatic differentiation, 40$\times$ faster than traditional numerical differentiation approaches in optimizing a large model, which we conclusively established is the only feasible way to train hybrid SciML models.
\end{itemize}

To summarize, our PBE solver is both computationally efficient and differentiable.
These two features make the solver proficient at its purpose, not only at performing forward simulations, but also in solving inverse problems, even before integrating any machine-learning components.
Furthermore, our solver lays a strong foundation for applying SciML techniques to PBEs in future work, including hybrid models.

\begin{acknowledgement}
M.A. would like to thank Mr. Yousuf Zaman and Mr. Oleksandr Prykhodko for the many fruitful discussions on parameter estimation and optimization behaviour, respectively.
\end{acknowledgement}

\appendix
\setcounter{table}{0}
\setcounter{equation}{0}
\setcounter{figure}{0}
\renewcommand{\thesection}{Appendix~\Alph{section}}
\renewcommand{\thefigure}{\Alph{section}.\arabic{figure}}
\renewcommand{\thetable}{\Alph{section}.\arabic{table}}
\renewcommand{\theequation}{\Alph{section}.\arabic{equation}}

\section{Extended Model Formulation}
\label{app-model}

The remaining definitions required to fully describe the crystallization model we presented in \seref{sec-model} include the solubility and the growth rates.
For the solubility expression, we use the following exponential by default
\begin{equation}
    c^*(T) = a \exp(bT)
    \label{eq-solubility_exponential}
  \end{equation}
where $a$ and $b$ are empirical constants, and $T$ is the temperature in degrees Celsius.
As for the growth rate, we default to the expression below
\begin{equation}
    G(S, T) = k_{\mathrm{g},m1}\exp(-\frac{k_{\mathrm{g},m2}}{T+273.15})(S-1)^{k_{\mathrm{g},m3}}
    \label{eq-growth2D_arrhenius}
\end{equation}
which only applies when the supersaturation $S$ exceeds one.
The index $m$ denotes the orthogonal length of the particles, and $G_m$ denotes the growth rate along those lengths.
The empirical parameters used for both \equsref{eq-solubility_exponential}{eq-growth2D_arrhenius} are shown in \tabref{tab-additional_parameters}.

It is worth noting that our solver framework is flexible in accepting any equation for the solubility and growth rates.
This design principle is what allowed us to change the growth rate expression to a polynomial in \seref{sec-pest}.

\begin{table}[ht!]
\caption{Additional simulation parameters used by default in all simulations unless stated otherwise}
\label{tab-additional_parameters}
\begin{tabular}{lll}
\hline
Name                            & Unit                    & Value         \\ \hline
\textit{Solubility parameters}  &                         &               \\
$a$                             & [\unit{\hour}]          & \num{3.37}    \\
$b$                             & [\unit{\degreeCelsius}] & \num{0.036}   \\
                                &                         &               \\
\textit{Growth rate parameters} &                         &               \\
$k_{\mathrm{g},11}$             & [\unit{\um\per\minute}] & \num{8.86e6}  \\
$k_{\mathrm{g},12}$             & [\unit{\K}]             & \num{2.45e3}  \\
$k_{\mathrm{g},13}$             & [-]                     & \num{3.7}     \\
$k_{\mathrm{g},21}$             & [\unit{\um\per\minute}] & \num{4.088e5} \\
$k_{\mathrm{g},22}$             & [\unit{\K}]             & \num{2.4e3}   \\
$k_{\mathrm{g},22}$             & [-]                     & \num{2.5}     \\ \hline
\end{tabular}
\end{table}

\section{Experimental Data Simulation}
\label{app-sim_experiments}

\begin{figure}
    \centering
    \includegraphics{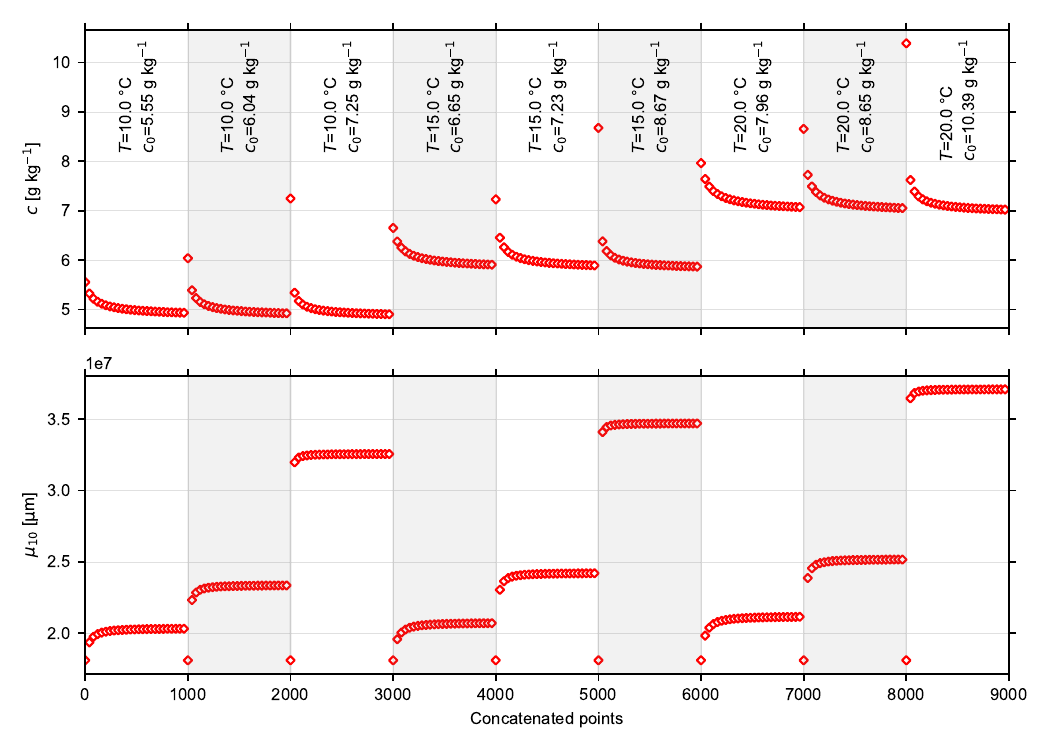}
    \caption{Concatenated time series of the experimental data. Each alternately shaded column represents a different experiment with a unique combination of temperature $T$ and initial supersaturation $S_0$. The latter is varied by varying the initial concentration $c_0$.}
    \label{fig-experimental_profiles}
\end{figure}

For optimizing the growth parameters from \seref{sec-pest}, we require experimental data, which we have generated \textit{in silico}.
We used largely the same simulation as the one outlined in \seref{sec-verify} (with parameters shown in \tabref{tab-default_parameters}).
To mimic an experimental campaign used to determine the growth rate model, we vary the temperatures and supersaturation from their default values shown in \tabref{tab-default_parameters}.
The temperatures used are [\qty{10}{\celsius}, \qty{15}{\celsius}, \qty{20}{\celsius}] and the initial supersaturations are [\num{1.15}, \num{1.25}, \num{1.5}], which leads to 9 unique combinations of temperature and supersaturation that form our experiments.
Note that to vary the initial supersaturation, we actually vary the initial concentration.
To calculate the initial concentration corresponding to the desired initial supersaturation, we use our solubility model (\equref{eq-solubility_exponential}).
Varying the supersaturation and temperature in this way is required to capture the dependence of the growth rate on these conditions in the proposed model.
It is important to mention that the growth rate model used to simulate experiments is \equref{eq-growth2D_arrhenius}, which is more realistic compared to the proposed model we fit to (\equref{eq-poly_growth_rate}).
The reason for this choice can be found in \seref{sec-pest}.
We then used the MOM to solve the simulation, and finally sampled concentration and crystal length at \num{600} uniform time intervals from the model solution to act as our true data (or $\mathbf{y}$, from \equref{eq-loss}).
The experimental data profiles are shown in \firef{fig-experimental_profiles}.

\section{Optimization Procedure}
\label{app-pest}

As mentioned in \seref{sec-pest}, the optimization algorithm that was used was Adam.
The Python library we used to implement Adam is Optax \cite{deepmind2020jax}, as it offers a flexible implementation of Adam that is compatible with JAX.
The initial learning rate used was 0.01, which is adaptive with successive iterations.
The other Adam hyperparameters, $\beta_1$ and $\beta_2$, were left at their default values of 0.9 and 0.999, respectively.
The optimizer was programmed to run for 100 iterations, with no convergence condition.
This was an intentional choice, as we weren't concerned with the final optimized result, but rather the average speed the optimizer takes towards convergence at each step.
Finally, the only constraint on the parameters being estimated is that they are non-negative.

As for the loss function itself, the residual sum of squares was selected, which minimizes the model prediction of the 9 experiments from \seref{app-sim_experiments}.
Most of the simulation parameters from \tabref{tab-default_parameters} still apply, with only two notable differences. First, bin length and width are both increased to \qty{5}{\um}, which reduces the time taken for the parameter estimation (but keeping relative trends the same). 
Second, the cooling temperature and initial concentration are changed to match each experiment shown in \firef{fig-experimental_profiles}.
However, we note that the choice of loss function is not central to our study, and that JAX's speed and differentiability improvements should be independent of the loss function used.

Anecdotally, we also compared the convergence of the optimizer using the two gradient algorithms, AD and ND from \seref{sec-pest}, and found no difference in most cases (when the numerical step size is within a sensible range).
The reason for this is obvious in hindsight, as Adam employs gradient normalization, which makes it insensitive to inaccuracies in the gradients calculated using ND.
For optimization algorithms that depend on the magnitude of the gradient, we expect that the exact value obtained by AD to be an added advantage over ND.
Unfortunately, we are unable to prove this statement within the scope of our study.

\begin{suppinfo}
Additional equations required to recreate our simulations are listed in \serefSI{si_sec-model}.
Further information on automatic differentiation, along with a simple example, are provided in \serefSI{si_sec-AD}.
Auxillary details on the computational methods used can be found in \serefSI{si_sec-comp_methods}.
\end{suppinfo}

\putbib
\end{bibunit}

\newpage
\nolinenumbers
\setlength\parindent{0pt}
\textbf{\Large{For Table of Contents Use Only}}
\bigskip

\textbf{Title:} Modern, Efficient, and Differentiable Transport Equation Models using JAX: Applications to Population Balance Equations\\
\textbf{Authors:} Mohammed Alsubeihi, Arthur Jessop, Ben Moseley, Cláudio P. Fonte, Ashwin Kumar Rajagopalan\\
\textbf{Graphical TOC Entry:}
\begin{figure}[h]
    \centering
    \includegraphics{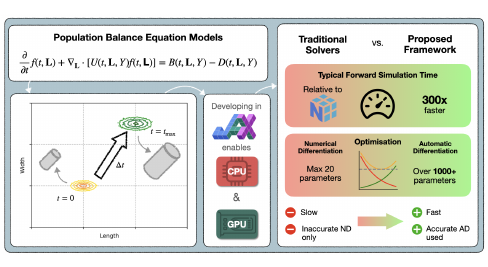}
\end{figure}

\textbf{Synopsis:}\\
A modern and efficient population balance equation solver is developed in JAX, a state-of-the-art programming library in Python. The solver enables faster, differentiable simulations, paving the way for hybrid scientific machine learning by a tight integration of neural networks with physics-based models.


\renewcommand{\thefootnote}{\fnsymbol{footnote}}
\renewcommand*\firef[1]{Figure~\ref{#1}}
\renewcommand{\thefigure}{S\arabic{figure}}
\setcounter{figure}{0}
\renewcommand*\tabref[1]{Table~\ref{#1}}
\renewcommand{\thetable}{S\arabic{table}}
\setcounter{table}{0}
\renewcommand{\thesection}{S\arabic{section}}
\setcounter{section}{0}
\renewcommand*\seref[1]{Section~\ref{#1} in the main text}
\renewcommand*\sesref[2]{Sections~\ref{#1}~and~\ref{#2} in the main text }
\renewcommand*\serefSI[1]{Section~\ref{#1}}
\renewcommand*\appendixref[1]{\ref{#1} in the main text}
\renewcommand*\equref[1]{eq~\ref{#1} of the main text}

\setcounter{equation}{0}
\renewcommand{\theequation}{\arabic{equation}}

\newpage

\singlespacing
\setlength\parindent{17pt}
\setcounter{page}{1}
\renewcommand{\thepage}{S\arabic{page}}

\newcommand{\secfnt}{\fontsize{14}{17}}
\newcommand{\ssecfnt}{\fontsize{12}{14}}
\titleformat{\section}
{\normalfont\secfnt\bfseries}{\thesection}{1em}{}
\titleformat{\subsection}
{\normalfont\ssecfnt\bfseries}{\thesubsection}{1em}{}

\begin{bibunit}

\nolinenumbers
\begin{center}
\textbf{\large{Supporting Information (SI) for ``Modern, Efficient, and Differentiable Transport Equation Models using JAX: Applications to Population Balance Equations"}}
\end{center}

\section{Additional Information on Model Formulation and Solution}

\label{si_sec-model}

\subsection{Discretization and Solution}
\label{si_sec-fvm}

To calculate the flux limiter $\phi$ the local smoothness $\theta$ must be calculated using
\begin{equation}
  \theta^n_{i-\frac{1}{2}} = \frac{f^n_{i-1} - f^n_{i-2}}{f^n_i - f^n_{i-1}}
\end{equation}
where $f^n_i$ is the value of the PSSD at time step $n$ and bin $i$. The smoothness is then used to calculate the flux limiter using the van Leer limiter function, given by
\begin{equation}
  \phi(\theta) = \frac{\theta + \abs{\theta}}{1 + \abs{\theta}}.
\end{equation}
The CFL condition, employed in the 2D algorithm, is used to determine the stable time step for each iteration and is given by
\begin{equation}
   \Delta t = \nu \, \min \left( \frac{\Delta L_1}{G_1}, \frac{\Delta L_2}{G_2} \right)
\end{equation}
where $\nu$ is the Courant number (which has been set to a value of 0.9), $\Delta t$ is the stable time step, $\Gamma_1$ and $\Gamma_2$ are the growth/dissolution rates along the two characteristic lengths, and $\Delta L_1$ and $\Delta L_2$ are the grid dimensions along the two lengths.

\subsection{Method of Moments}
\label{si_sec-mom}

The method of moments can be derived for the case of well-mixed batch crystallizer. For particles approximated using two dimensions, which are subject to growth, a set of ODEs to describe the evolution of the moments over time can be given by
\begin{equation}
    \dv{\mu_{p,q}}{t} = pG_1\mu_{p-1,q} + qG_2\mu_{p,q-1}
    \label{eq-mom2D}
\end{equation}
where $\mu_{p,q}$ is the cross moment of order $p$ and $q$, and $G$ is the growth rate in each characteristic length. This system is simple to solve using an off-the-shelf ODE solver, provided the moments of the initial distribution are known. To directly compute the $pq$ cross moment of a 2D PSSD, the following equation is used
\begin{equation}
  \mu_{p,q} = \int_{0}^{\infty}\int_{0}^{\infty}L_1^pL_2^qf(L_1,L_2,t)\mathrm{d}L_1\mathrm{d}L_2 
  \quad \forall p,q \in \mathbb{N} \label{eq-moment2D}.
\end{equation}

\subsection{Initial Distribution Generation}

Inputting any measured distribution is certainly possible within our framework to simulate any experimental distribution. 
However, since this is a computational study, an initial, or seed, distribution is usually generated \textit{in silico} instead to effectively investigate how the solver responds to different distributions. 
The seed is assumed to be representatively sampled, and the sample characterized to obtain the PSSD of the sample $f_\mathrm{sample}$, which is common in experimental workflows.
This sample PSSD can also be assumed to follow a statistical distribution, such a normal, or log-normal distribution, which can be easily calculated based on a few given parameters, including the mean and the covariance of the distribution.
Our framework supports normal and log-normal distributions for the seed PSSD, though this is trivial to extend to any similarly generated distribution.
To scale up the sample PSSD to the number density of the crystallizer, the mass of the seed $m_{0}$, which is provided as an initial condition, can be used to calculate a scale factor.
The scale factor $\beta$ can hence be calculated using
\begin{equation}
  \beta = \frac{f_0}{f_\mathrm{sample}} = \frac{m_0}{m_\mathrm{sample}} = \frac{m_\mathrm{seed}}{\rho_\mathrm{c} k_\mathrm{v} \int_0^\infty \int_0^\infty L_1 L_2^2 f_\mathrm{sample} \mathrm{d}L_1 \mathrm{d}L_2}
\end{equation}
where the mass of the sample $m_\mathrm{sample}$ is obtained by integrating the PSSD of the sample.

\section{Additional Information on Automatic Differentiation}
\label{si_sec-AD}

\subsection{Fundamentals of Automatic Differentiation}
\label{si_sec-autodiff_basics}

We provide an overview of automatic differentiation, which we needed to understand its applicability to PBEs. To begin, the alternatives to AD are discussed.

One alternative to AD is symbolic differentiation, which yields an exact expression for the derivative, essential for understanding and interpreting rates of change.
However, applying symbolic differentiation is undesirable in many situations, as it is often tedious and computationally inefficient, especially with the complex functions common in science and engineering.
Also, in many cases, the expression for the derivative is not important, only the value it yields.
In those cases, numerical differentiation is commonly used, which is more flexible when compared to symbolic differentiation, but suffers from truncation error.
Numerical differentiation also does not scale well to Jacobians of multidimensional functions of the form $\mathbb{R}^p\to\mathbb{R}^q$, as $2pq$ evaluations are required at minimum to obtain the Jacobian of the function with the simplest possible scheme.
This also applies to symbolic differentiation, as each element of the Jacobian must be derived separately.

\begin{figure}[htbp!]
    \centering
    \includegraphics{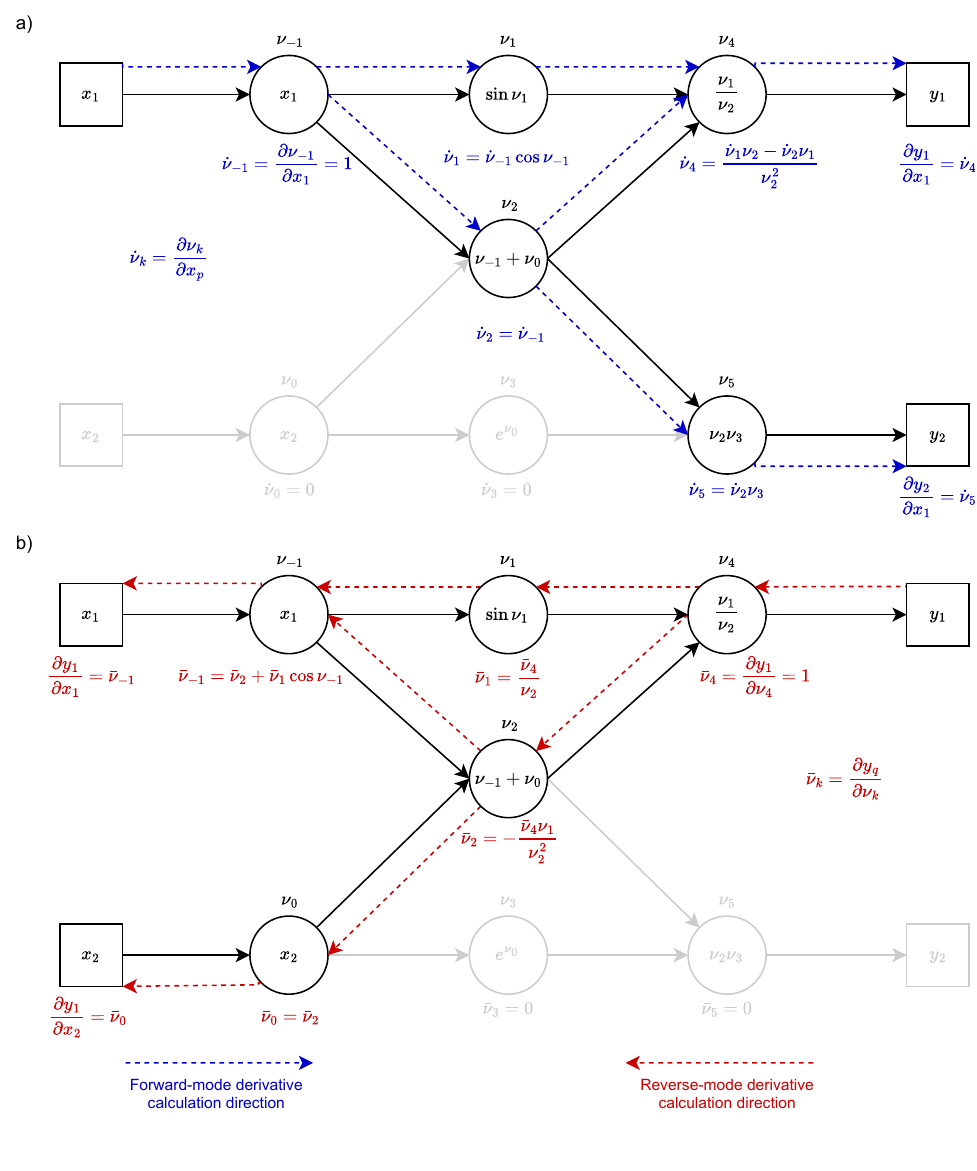}
    \caption[Computational graph showing automatic differentiation applied to an example function.]{Computational graph of the function $y_1 = \frac{sinx_1}{x_1+x_2}$, $y_2 = (x_1+x_2)e^{x_2}$, in which the function is decomposed to primitive operations on intermediates, $\nu_k$. a) Forward-mode determines the derivatives of all outputs with respect to one input, $x_1$, in a single forward pass. b) Reverse-mode determines the derivatives of one output, $y_1$, with respect to all inputs, in two passes. First, a forward pass is used to evaluate and store primitives, which a backwards pass then uses to calculate the derivatives.}
    \label{fig-AD_example}
\end{figure}

AD is distinct from both aforementioned differentiation techniques, as it can calculate exact (to machine precision) values of derivatives efficiently, and can be applied to any mathematical program \cite{baydinAutomaticDifferentiationMachine2018}.
In both modes, every expression is broken down into primitive operations (e.g. addition, exponentiation) on intermediate values (known as primitives), an example of which is shown in \firef{fig-AD_example}.
In forward-mode, shown in panel a) of \firef{fig-AD_example}, each primitive is augmented with its corresponding derivative (called a tangent).
These tangents are obtained through symbolic differentiation of the primitive operations, and then propagated using the chain rule to obtain the derivative of all outputs with respect to a single input.
Reverse-mode is similar, but instead, quantities called adjoints are propagated `backwards' through the computation tree, as shown in panel b) of \firef{fig-AD_example}.
Forward-mode and reverse-mode calculate derivatives with the same computational efficiency, and are particularly powerful when applied to multidimensional functions of the form $\mathbb{R}^p\to\mathbb{R}^q$.
The forward-mode can determine the gradient of all function outputs with respect to one input in a single pass, whereas the reverse-mode does so for one output with respect to all inputs.
Therefore, the forward-mode is much more efficient when the function outputs significantly outnumber the function inputs, i.e., $p<<q$ (tall Jacobian), and vice versa for the reverse-mode, which is more efficient given $p>>q$ (wide Jacobian).
This is what makes reverse-mode particularly suited to inverse problems involving scalar problems, which are common in optimization, and especially ML.
For example, the gradient with respect to the (very commonly) millions of model weights in a neural network can be obtained in a single backward pass, which is precisely what backpropagation does.

Computing the derivative using AD is provably efficient through the `cheap gradient principle' \cite{griewankFundamentalsForwardReverse2008}, which states that computing one column of the Jacobian using forward-mode, or one row using reverse-mode, is often as fast as the computation of the function itself, and at maximum AD takes 6 times longer.
However, the main drawback of reverse-mode is the increased memory usage, although this can be mitigated with the use of memory checkpoints, a technique which JAX offers convenient support for.
Reverse- and forward-mode can also be combined to obtain higher-order derivatives, such as the computation of the Hessian matrix.
The differentiability of our model means we can obtain the derivative of any output involving our model with respect to any of its inputs.
This is especially powerful in dealing with inverse problems, where a scalar objective function is minimized subject to a vector of model inputs, and also provides a strong foundation for SciML techniques, particularly hybrid modelling.
Obtaining the exact derivative using JAX is as easy as applying \verb|grad()| to the function of interest, with many other functions available for finer control, including which mode to use, and more.

\subsection{Full Example of Reverse and Forward Modes}
Here we show an example of automatic differentiation applied to the function presented in \serefSI{si_sec-autodiff_basics}.
First, we illustrate forward mode applied to obtain the derivatives of both outputs with respect to $x_1$, in the same way as panel~a) of \firef{fig-AD_example}.
The input values used in the example are: $x_1=5$, $x_2=10$.
We summarize the forward-mode example in \tabref{tab-auto_diff_forward_example} and the reverse-mode example in \tabref{tab-auto_diff_reverse_example}.
\begin{table}[h]
  \centering
  \caption[Forward-mode AD example.]{Example of forward mode applied to the simple function introduced in \serefSI{si_sec-autodiff_basics}, with input values of $x_1=0.5$ and $x_2=1$. Each primitive calculation is followed by a tangent calculation, which are obtained together in a single pass. The final result is the derivative of every output with respect a single output, $x_1$.}
  \label{tab-auto_diff_forward_example}
  \begin{tabular}{@{}ll@{}}
  \toprule
  Primal                 & Tangent                                         \\ \midrule
  $\nu_{-1}=x_1=0.5$     & $\dot{\nu}_0=1$                                 \\
  $\nu_0=x_2=1$          & $\dot{\nu}_0=0$                                 \\
  $\nu_1=\sin\nu_1=0.48$ & $\dot{\nu}_1 = \dot{\nu}_{-1}\cos\nu_{-1}=0.88$ \\
  $\nu_2=\nu_{-1}=1.5$   & $\dot{\nu}_2=\dot{\nu}_{-1}=1$                  \\
  $\nu_3=e^{\nu_0}=2.72$ & $\dot{\nu}_3 = 0$                               \\
  $\nu_4=\frac{\nu_1}{\nu_2}=0.32$ & $\dot{\nu}_4=\frac{\dot{\nu}_1\nu_2-\dot{\nu}_2\nu_1}{\nu_2^2}=0.37=\frac{\partial y_1}{\partial x_1}$ \\
  $\nu_5=\nu_2\nu_3=4.08$          & $\dot{\nu}_{5} =\dot{\nu}_2\nu_3=2.72=\frac{\partial y_2}{\partial x_1}$                               \\ \bottomrule
  \end{tabular}
\end{table}
\begin{table}[h]
  \centering
  \caption[Reverse-mode AD example.]{Example of reverse mode applied to the simple function introduced in \serefSI{si_sec-autodiff_basics}, with input values of $x_1=0.5$ and $x_2=1$. First, the primitives are calculated and stored in a forward pass. Then, adjoints are propagated using the chain rule in a backwards pass. The final result is the derivative of one output, $y_1$, with respect to every input.}
  \label{tab-auto_diff_reverse_example}
  \begin{tabular}{@{}ll@{}}
  \toprule
  Primal Calculation (Forward Pass) & Adjoint Calculation (Backwards Pass)                  \\ \midrule
  $\nu_{-1}=x_1=0.5$                & $\bar{\nu}_5=0$                                       \\
  $\nu_0=x_2=1$                     & $\bar{\nu}_4=1$                                       \\
  $\nu_1=\sin\nu_1=0.48$            & $\bar{\nu}_3 = 0$                                     \\
  $\nu_2=\nu_{-1}=1.5$              & $\bar{\nu}_2=-\frac{\bar{\nu}_4\nu_1}{\nu_2^2}=-0.21$ \\
  $\nu_3=e^{\nu_0}=2.72$            & $\bar{\nu}_1 = \frac{\bar{\nu}_4}{\nu_2}=0.67$        \\
  $\nu_4=\frac{\nu_1}{\nu_2}=0.32$ & $\bar{\nu}_0=\bar{\nu}_2=-0.21=\frac{\partial y_1}{\partial x_2}$                            \\
  $\nu_5=\nu_2\nu_3=4.08$          & $\bar{\nu}_{-1} =\bar{\nu}_2+\bar{\nu}_1\cos\nu_{-1}=0.37=\frac{\partial y_1}{\partial x_1}$ \\ \bottomrule
  \end{tabular}
  \end{table}

We can verify that we have completed forward and backwards mode correctly by evaluating the Jacobian of the function using symbolic differentiation thusly
\begin{equation*}
  \mathbf{J} = 
  \begin{bmatrix}
    \pdv{y_1}{x_1} & \pdv{y_1}{x_2} \\
    \pdv{y_2}{x_1} & \pdv{y_2}{x_2}
  \end{bmatrix} =
  \begin{bmatrix}
    \frac{(x_1 + x_2)\cos x_1 - \sin x_1}{(x_1 + x_2)^2} & -\frac{\sin x_1}{(x_1 + x_2)^2} \\
    e^{x_2} & (x_1 + x_2 + 1)e^{x_2}
  \end{bmatrix}
\end{equation*}

By substituting a value of $x_1=0.5$ and $x_2=1$, we obtain the following Jacobian
\begin{equation*}
  \mathbf{J} =
  \begin{bmatrix}
    0.37 & -0.21 \\
    2.72 & 6.80
  \end{bmatrix}
\end{equation*}
which is a match to our forward-mode (first column), and reverse-mode (first row), calculations.

\section{Additional Information on Computational Solvers Comparison}
\label{si_sec-comp_methods}

\subsection{Other Languages and Libraries}

The alternatives we compare our JAX solver to are the NumPy, C++, and CUDA solvers.
To ensure a fair comparison, the algorithms across the different solvers we kept as close as possible.
All solvers use floating point numbers of double precision (64-bit) to ensure accuracy and consistency.
The NumPy solver is the closest to our JAX solver, since they are both Python libraries, and JAX uses a NumPy-like interface.
The only notable difference is that the NumPy arrays are mutable, meaning they can be updated in-place in memory, whereas JAX arrays are immutable, which means that a copy is created instead for every array that is modified. 
One would imagine this to be a huge issue for a solver such as the one developed in this work given that there are many arrays, all potentially very large, that need to be updated every single time step.
However, this anticipated massive copying overhead is all handled and optimized by the JIT compiler, which facilitates in place updates.
For the C++ solver, (row-major) loops are used instead of the vectorized operation available in both JAX and NumPy.
Finally, the CUDA solver is identical to the C++ one, except that it is compiled by \verb|nvcc|, and uses CUDA kernels to update the flux limiters, calculate the smoothness, and update the PSSD (horizontally and vertically separately using dimensional splitting).
Both the CUDA and C++ solvers are called from within Python using extension modules created using the pybind11 C++ library, with CMake used to automate building them. \verb|g++| and \verb|nvcc| are used to compile the C++ and CUDA solvers, respectively, with both using the \verb|-O3| flag for aggressive compiler optimizations.

\subsection{Simulation Verification}
\label{si_sec-sim_verification}



Verifying the simulations by individually plotting the different results and simulations for each case study, as we demonstrated in \firef{fig-concentration_moment_profiles} of the main text, quickly becomes infeasible.
This issue was made worse by the fact that the underlying code used to generate the simulations was constantly being improved, which required the simulations to be rerun.
Thus, we required an automated way of testing the results on a regular basis for any discrepancies, to maintain research integrity and for our own peace of mind.
Conveniently, many testing frameworks are available in Python to simplify this process.
One such framework is pytest, which we used to automate the testing of simulations in the main text.
To be specific, all simulations used in \seref{sec-comp_methods} were automatically tested to verify that all FVM solvers produced concentration and moment profiles that matched the method of moments to a prespecified tolerance.
In addition, the FVM solvers were compared to each other by verifying that they produce the same final PSSD.
The tests were regularly re-run (especially after large changes to the codebase, and especially before paper submission) to verify that the simulations were valid.
Pytest produces a report upon the completion of the testing procedure detailing which specific tests failed, if any.
This was used to diagnose any discrepancies, and eventually rectify them.

\putbib
\end{bibunit}

\end{document}